\documentclass[dvips]{article}

\usepackage{icrc2011}

\title{Multifrequency Variability and Correlations from Extensive Observing Campaigns of Mkn 421 and Mkn 501 in 2009}

\newcommand{\etal}{\MakeLowercase{\textit{et al. }}} 
\shorttitle{Barres de Almeida \etal Multifrequency Campaigns of Mkn 421 and Mkn 501 in 2009}

\authors{U. Barres de Almeida$^{1}$, D. Paneque$^{1}$, N. Nowak$^{1}$,  N. Strah$^{2}$ and D. Tescaro$^{3}$
on behalf of the Fermi-LAT, MAGIC and VERITAS collaborations and other groups involved in the 2009 MWL campaign.\\}
\afiliations{$^1$Max-Planck-Institut f\"{u}r Physik (Werner-Heisenberg-Institut), D-80805 M\"{u}nchen, Germany.\\ 
$^2$Technische Universit\"{a}t Dortmund, D-44221 Dortmund, Germany.\\
$^3$Universit\`{a} di Siena and INFN Pisa, I-53100 Siena, Italy.}
\email{ulisses@mppmu.mpg.de}

\abstract{We are peforming an unprecedentedly long and dense monitoring of the multifrequency (radio to TeV) emission 
from the classical TeV blazars Mkn 421 and Mkn 501. These objects are among the brightest X-ray/TeV blazars in the sky and 
among the few sources whose spectral energy distributions (SED) can be completely characterised by the current instruments.
This is a multi-year and multi-instrument programme which includes the participation of VLBA, Swift, RXTE, MAGIC, VERITAS,
Whipple, the Fermi/LAT Gamma-ray Observatory, GASP-WEBT, among other collaborations and instruments which combined provide
the most detailed temporal and energy coverage of these sources to date. In this proceedigns we will focus mostly on the 
results obtained with the multifrequency data from 2009, for which the SEDs of Mkn 421 and Mkn 501 are very similar and can
be described by a one-zone synchrotron self-Compton scenario. We will report on the 
multifrequency variability derived from these data.}

\begin{document}
\maketitle

\section{Introduction}

The northern-hemisphere high-energy peaked BL Lac (HBL), Mkn 421 \cite{Punch92} and Mkn 501 \cite{Quinn96} are the first 
extragalactic 
sources detected in the TeV range, and are two of the brightest and most active sources in the extragalactic VHE sky.
Due to their proximity ($z = 0.031$ and $z = 0.034$, respectively) and consequently high flux densities at very-high 
energies (VHE), which 
benefit from a low extragalactic background light (EBL) absorption level, an exquisite characterisation of the gamma-ray 
component of the spectrum is 
possible for these objects as for no other blazar. An excellent coverage of the VHE component of their 
SED is achievable with the current generation of Cherenkov telescopes (see e.g., \cite{MAGIC09a, MAGIC10a}) and 
the combined
activity of the Fermi Large Area Telescope (LAT) allows for an unprecedented over five decades energy 
coverage of the high-energy part of the SED \cite{Fermi11a, Fermi11b}.

During the flaring states, at which observations and multiwavelength campaigns usually concentrate, the studies have shown
that their time-averaged SED can be well characterised by synchrotron self-Compton (SSC) models, thus establishing a 
likely leptonic origin for the gamma-ray photons (e.g., \cite{Katarzynski10, Mastichiadis08}). The emission from
both these classical TeV blazars is characterised by variability on various timescales, from months down to minutes
\cite{MAGIC07, Zweerink97} and a general trend of correlation between the VHE and X-ray emission 
\cite{Fossati08}, though ``orphan'' gamma-ray flares as well as X-ray outbursts with no significant enhancement at higher
energies have also been documented (e.g., \cite{Tluczykont10}).

\section{Extensive Monitoring Campaigns in 2009}

Blazars emit radiation over a broad energy range, with flux and spectral variability being registered throughout the 
spectrum. This characteristic implies that an in-depth study of blazar physics requires closely
contemporaneous multi-frequency monitoring by a number of different instruments. Furthermore, due to the large amount
of telescope time required for good-coverage MWL follow-ups, datasets are heavily biased towards active 
states, and usually concentrate on extreme flaring episodes, which introduces an unavoidably skewed view of the jet 
physics. Somehow surprisingly, despite the years of study and observational efforts, intense, unbiased and long-term 
full-SED MWL monitoring of blazars lacked.


\begin{figure}[htbp]
\begin{center}
\includegraphics[angle=0, width=0.4\textwidth]{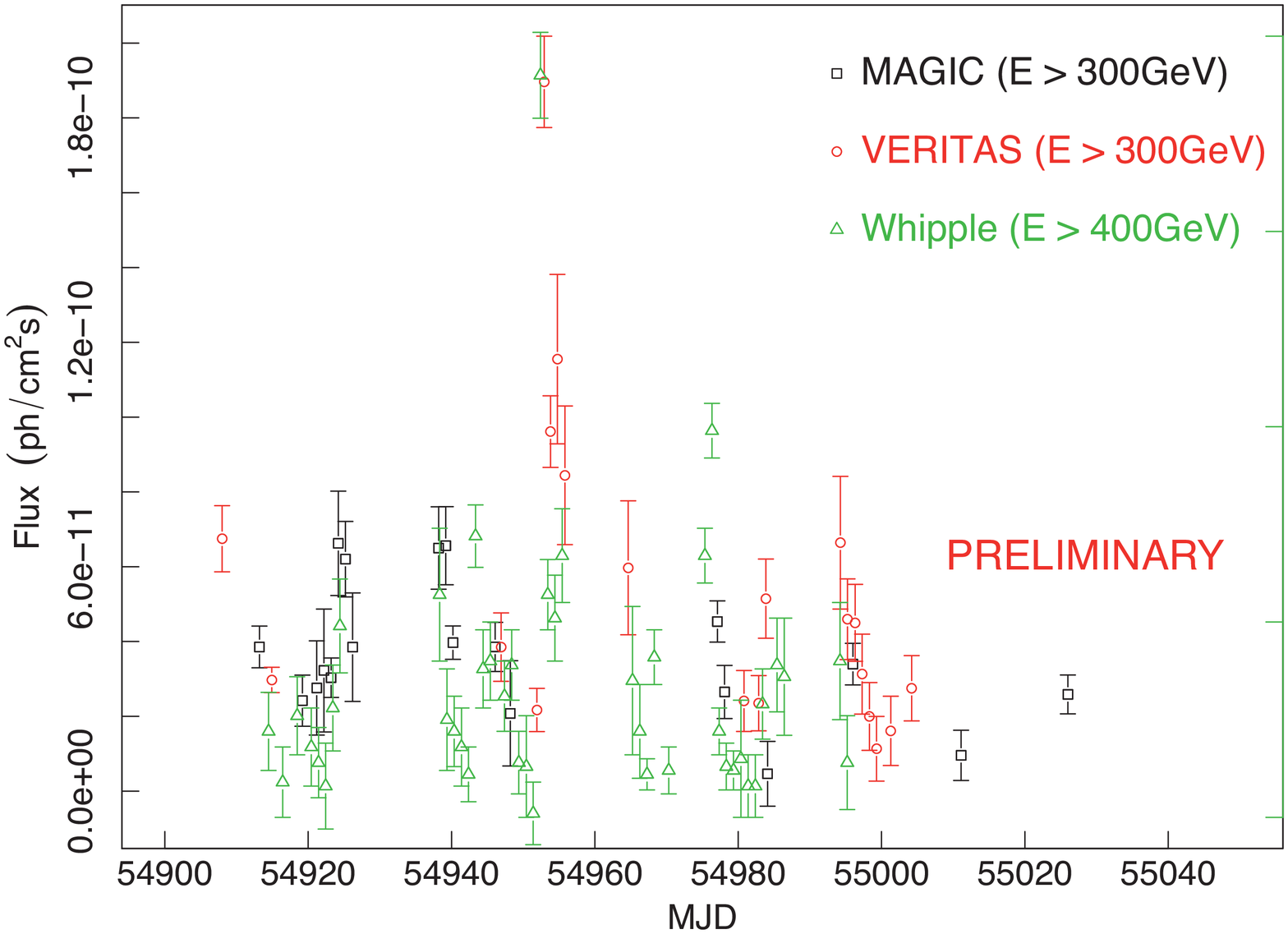}
\includegraphics[angle=0, width=0.4\textwidth]{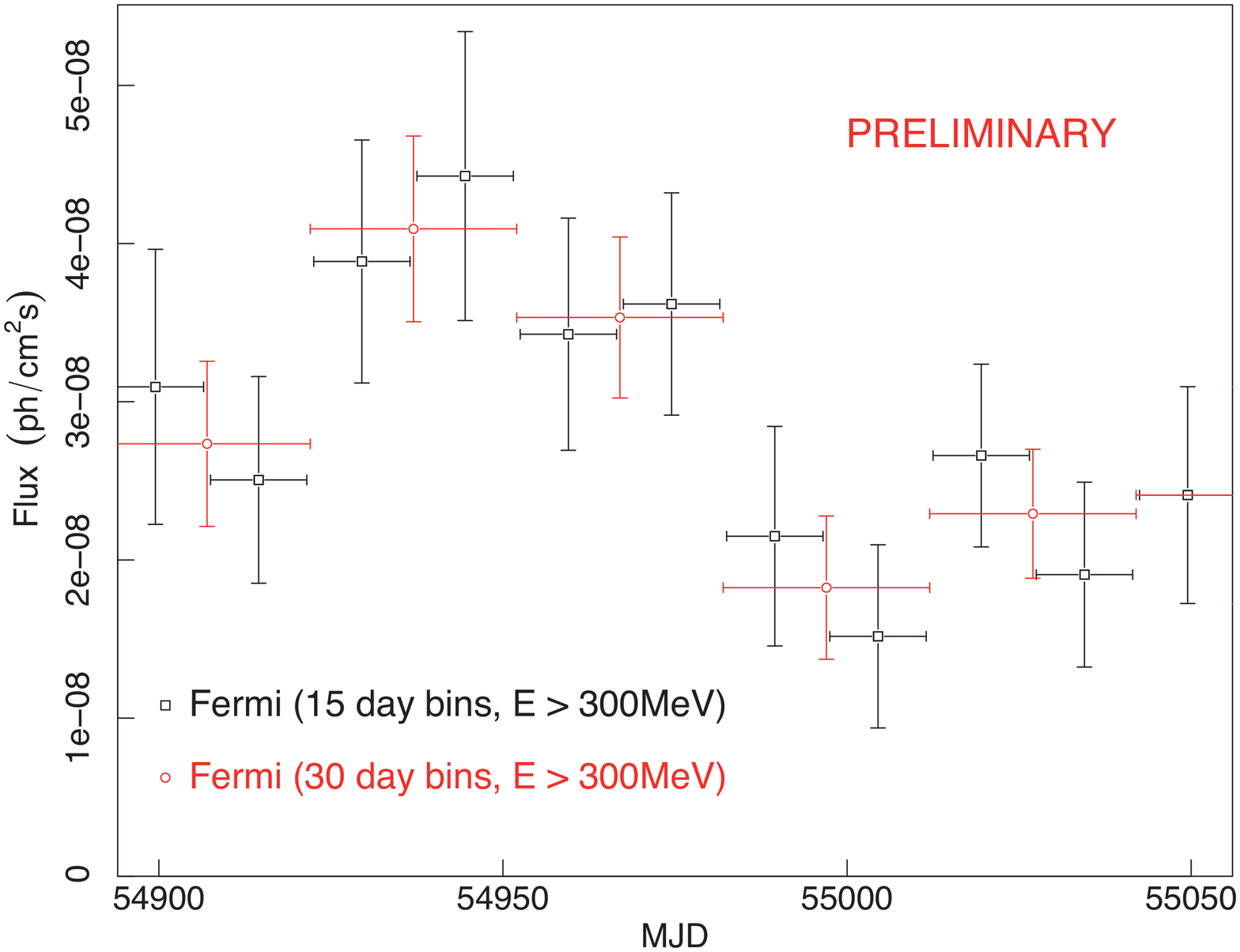}
\includegraphics[angle=270, width=0.45\textwidth]{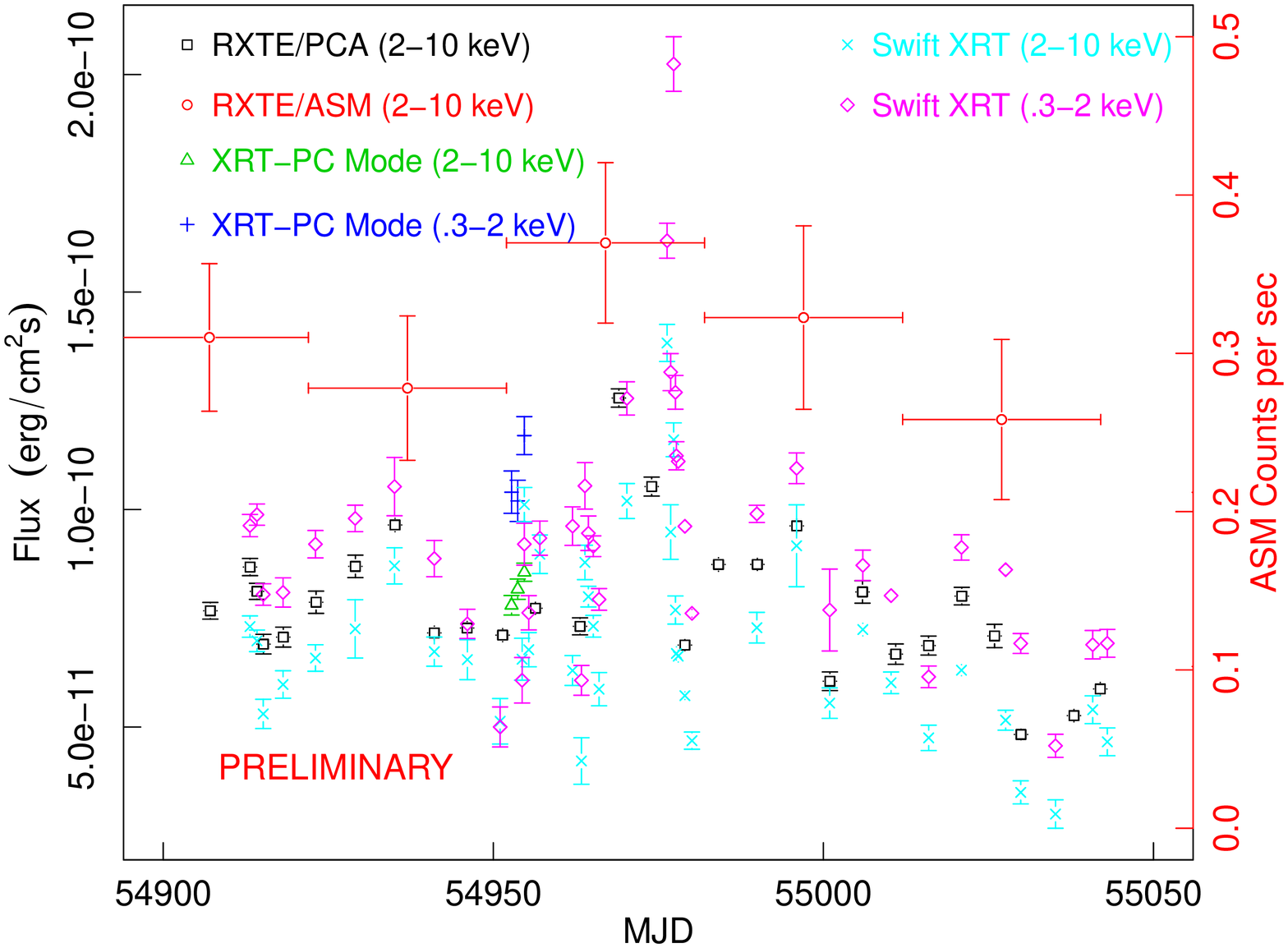}
\includegraphics[angle=0, width=0.4\textwidth]{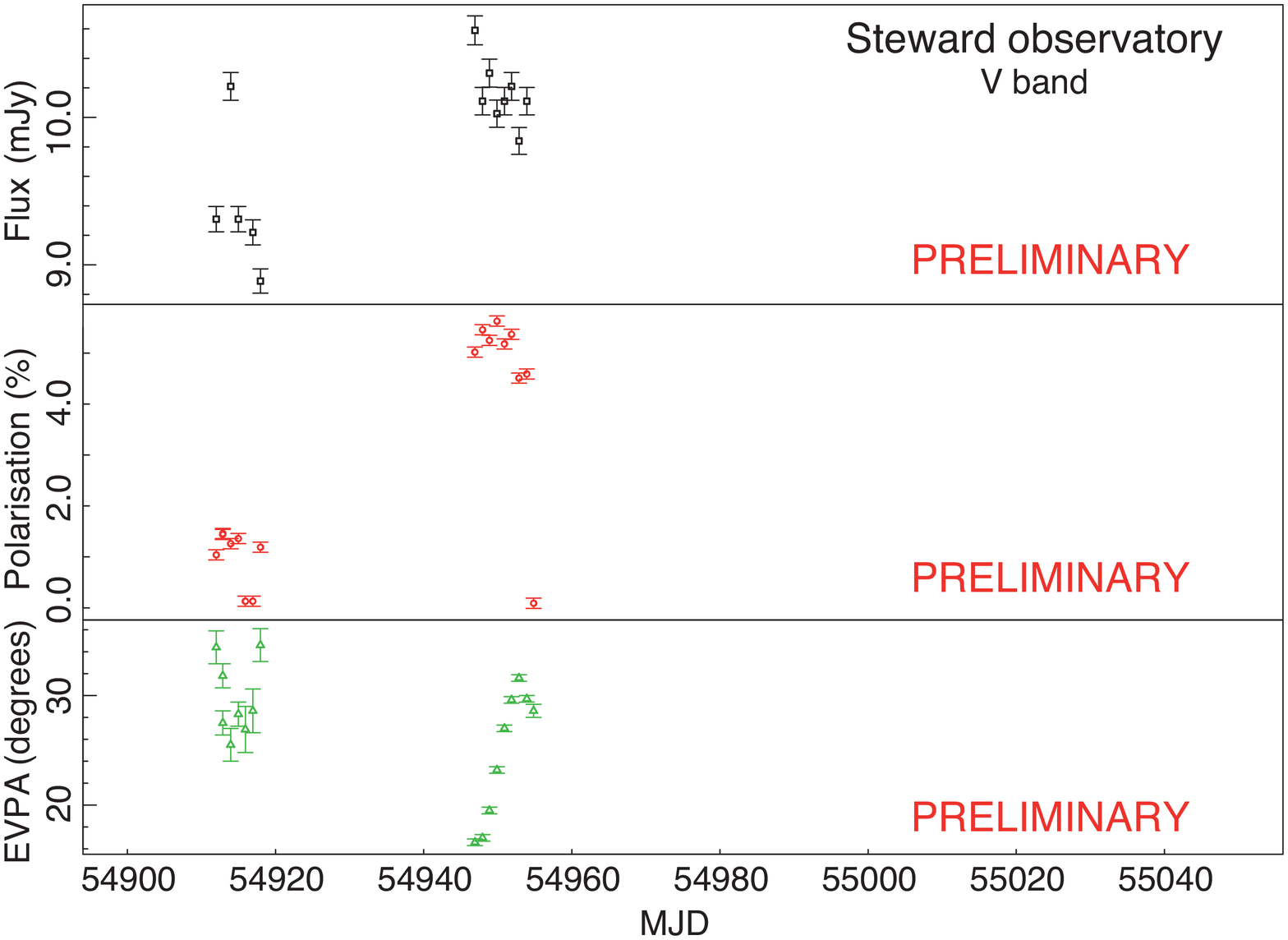}
\caption[]
        {{\it Top to bottom panel:}VHE (MAGIC, Whipple, VERITAS), Fermi-LAT, X-rays (RXTE and Swift) and 
          Steward Optical polarimetry lightcurves of Mkn 501 between MJDs 54900 and 55050. 
          Notice the flare at MJD 54950, and abscence of X-ray and Fermi activity. 
          The activity in polarisation coincides with the time of the flare.
        }
\label{fig:Mkn501_lc}
\end{center}
\end{figure}

In 2009, a series of extensive multi-instrument campaigns to monitor the two bright HBLs Mkn 421 and 501 was started (and 
is still ongoing) in an attempt to bridge this long-standing gap and bring about a more fiducial view of the physics of
AGN jets\footnote{Extensive details on these campaigns can be found at: \\ 
\url{https://confluence.slac.stanford.edu/GLAMCOG/Campaign+on+Mrk421}\\
\url{https://confluence.slac.stanford.edu/GLAMCOG/Campaign+on+Mrk501}}.
Both campaigns were conducted for 4.5 months and counted with quasi-daily observations of over 25 instruments covering
simultaneously the entire spectral range of the SED, from cm radio wavelengths to TeV gamma-rays. A particularly unique and
important aspect of the campaigns is that the sources were followed during the entire period and with equally intense 
coverage, regardless of their activity state.
 

\subsection{Broadband SEDs of Mkn 421 and Mkn 501}

Results on the broadband, time-averaged SEDs obtained from the entire campaign for Mkn 421 and Mkn 501, including a 
detailed discussion of the modelling of the respective SED, can be found in two recent 
publications \cite{Fermi11a, Fermi11b} and are only briefly mentioned.


Both sources were found to be in a low state throughout the campaign, without any major flares, except for Mkn 501 
during three days of observations, when an ``orphan'' TeV outburst was detected. For both objects, this was the first 
time that 
simultaneous observations by the Fermi-LAT and the VHE instruments MAGIC and VERITAS allowed for a complete coverage of the
high-energy component of the SED (the inverse-Compton bump) for over five orders of magnitude range in energy. This 
allowed a detailed characterisation of the quiescent SED of these important HBLs.

Profiting from this excellent spectral data, the SED was modelled using a one-zone SSC model which was found to explain
very well the entire set of observations. A characteristic of these fits is the similarity 
between the physical parameters obtained for both sources, such as the values of basic physical parameters of the emitting
region like size of the emitting blob and the magnetic field. The Doppler factors were shown to be of the same order of 
magnitude, and for both objects the modelled electron population shared equally very similar properties. The 
characteristics
of the electron populations is probably resulting from common properties of the acceleration and cooling mechanisms at the
source and were argued to point to some common property of the jets and the acceleration process in these two blazars, 
.


\subsection{Multiband Lightcurves}

In this paper are presented the preliminary lightcurve and variability results from the two campaigns. 
Over 25 instruments took part in the observations, generating a wealth of data that cannot be shown
here in its entirety.

\subsubsection{Markarian 501}
Figure \ref{fig:Mkn501_lc} (top panel) shows the VHE lightcurves for Mkn 501 taken with MAGIC, VERITAS and Whipple. The 
dataset is dominated by the presence of a flare lasting for a few days around MJD 54950. The Whipple data (with energy 
threshold of 400 GeV) is presented here 
normalised to a threshold of 300 GeV according to a power law spectrum with index -2.5, to conform with the other 
instruments' characteristics. This VHE flare, which seems not to have any counterparts in the X-rays 
(a so-called ``orphan flare''), is nevertheless correlated with a 
5\% increase in the optical polarisation of the source as measured at the Steward observatory 
(see \cite{Pichet} for the details on this event). 
This increase is considered quite significant given that at low states the typical polarisation levels 
of Mkn 501 are around 1-3\%. Accompanying this polarisation flare, a rotation of the electric vector position angle (EVPA) 
by 15$^\circ$ was seen. Despite the limited time coverage of the source in polarisation, the rotation profile is well 
sampled, showing a reversed trend in the last few days. Such rotation events can originate from a number of reasons, such
as episodic re-ordering of the magnetic field at the emitting region, due for example to shocks or injection of freshly 
accelerated particle populations, as well as movements of the plasma through large scale ordered fields, all of which could
have a strict connection with the observed VHE flare.  


\subsubsection{Markarian 421}  
\begin{figure}[htbp]
\begin{center}
\includegraphics[angle=0, width=0.4\textwidth]{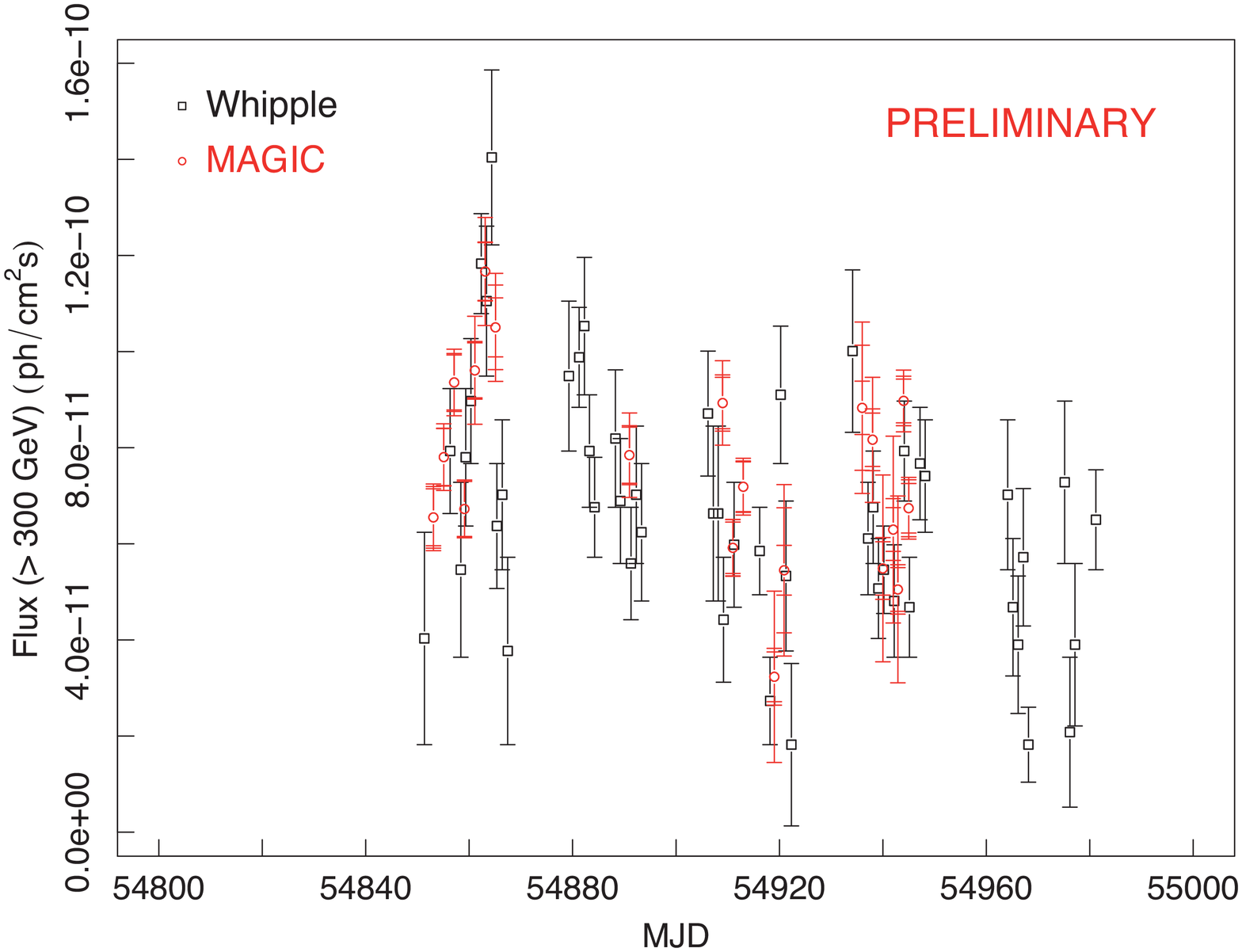}
\includegraphics[angle=0, width=0.4\textwidth]{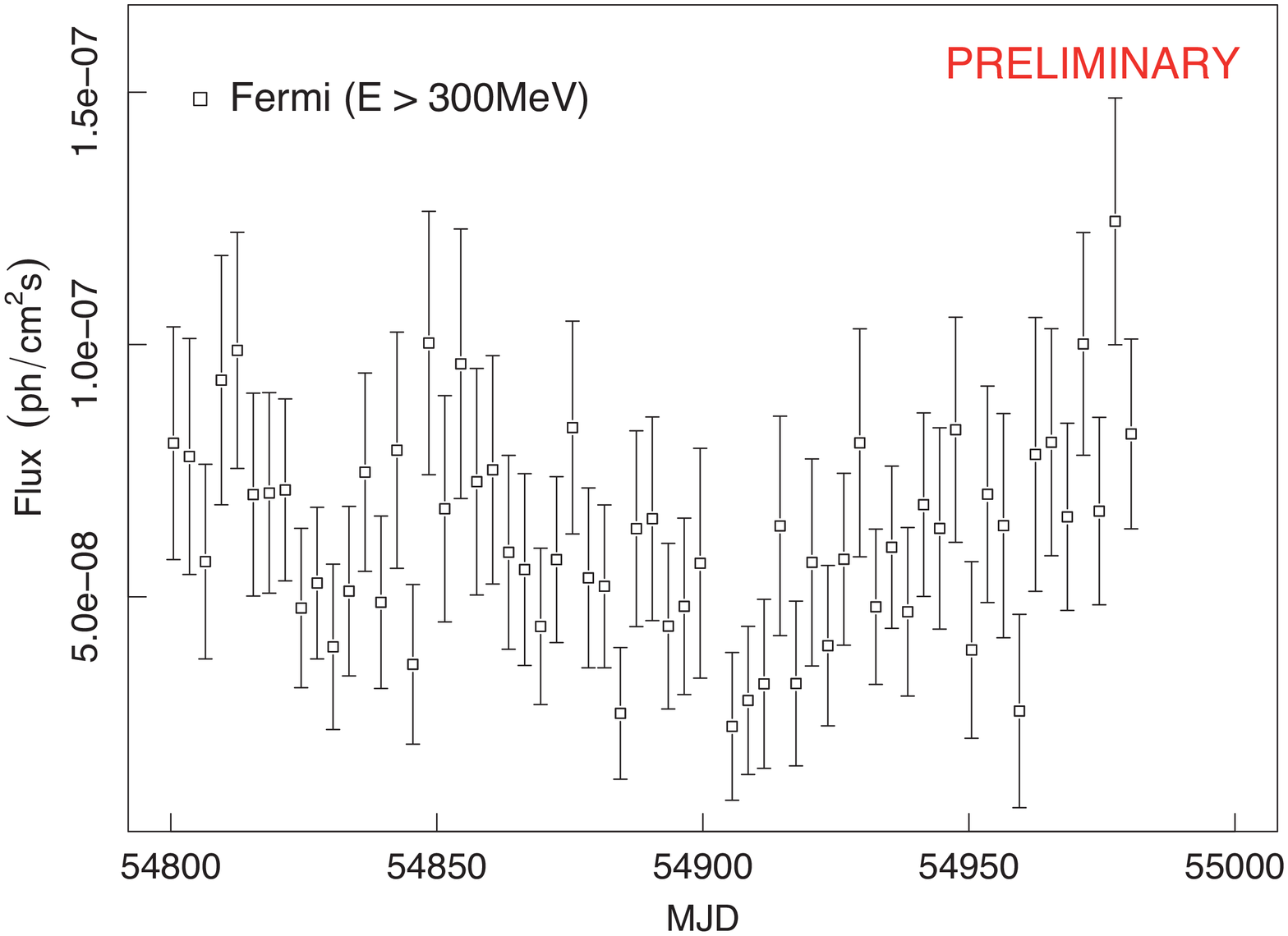}
\includegraphics[angle=270, width=0.45\textwidth]{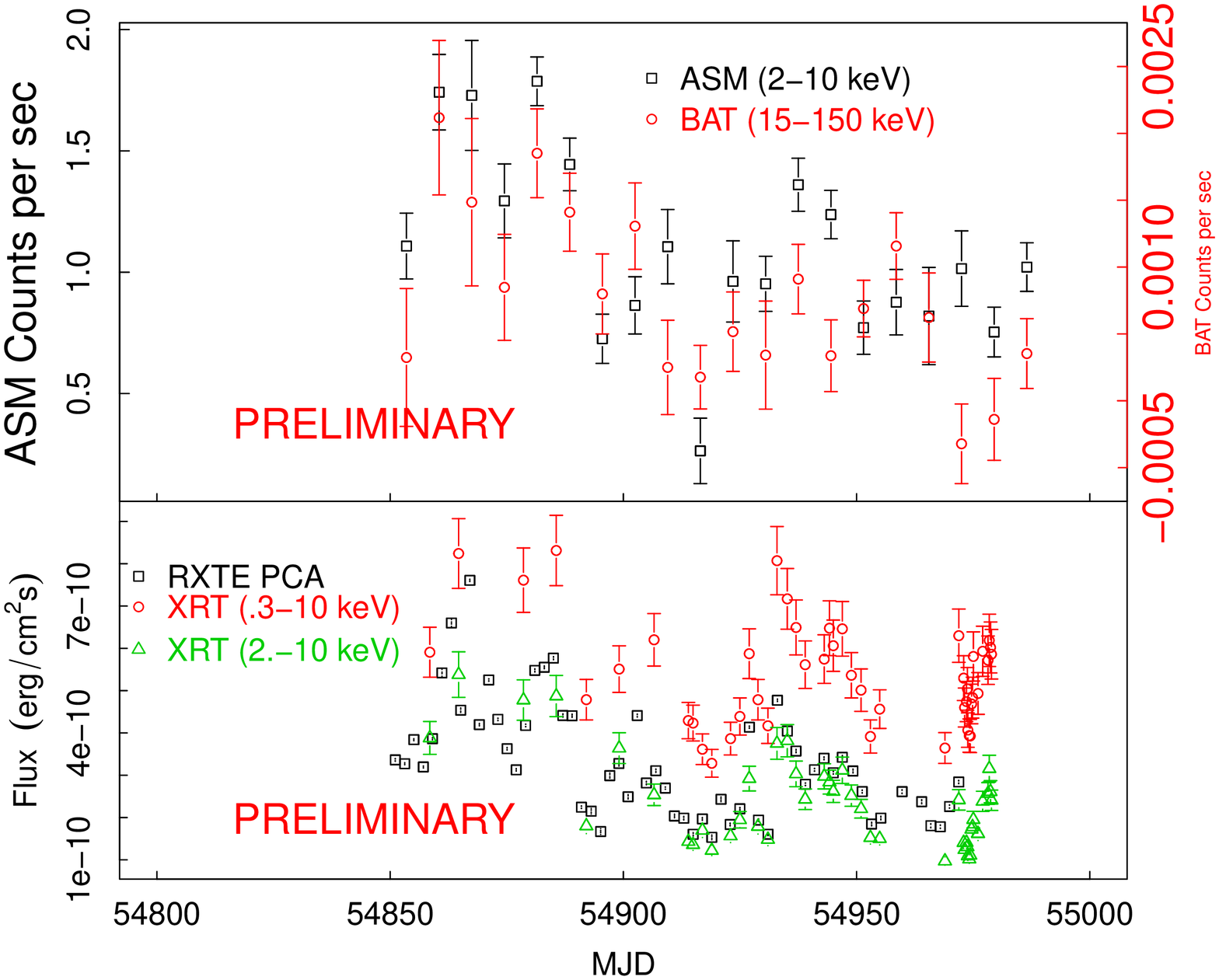}
\includegraphics[angle=0, width=0.4\textwidth]{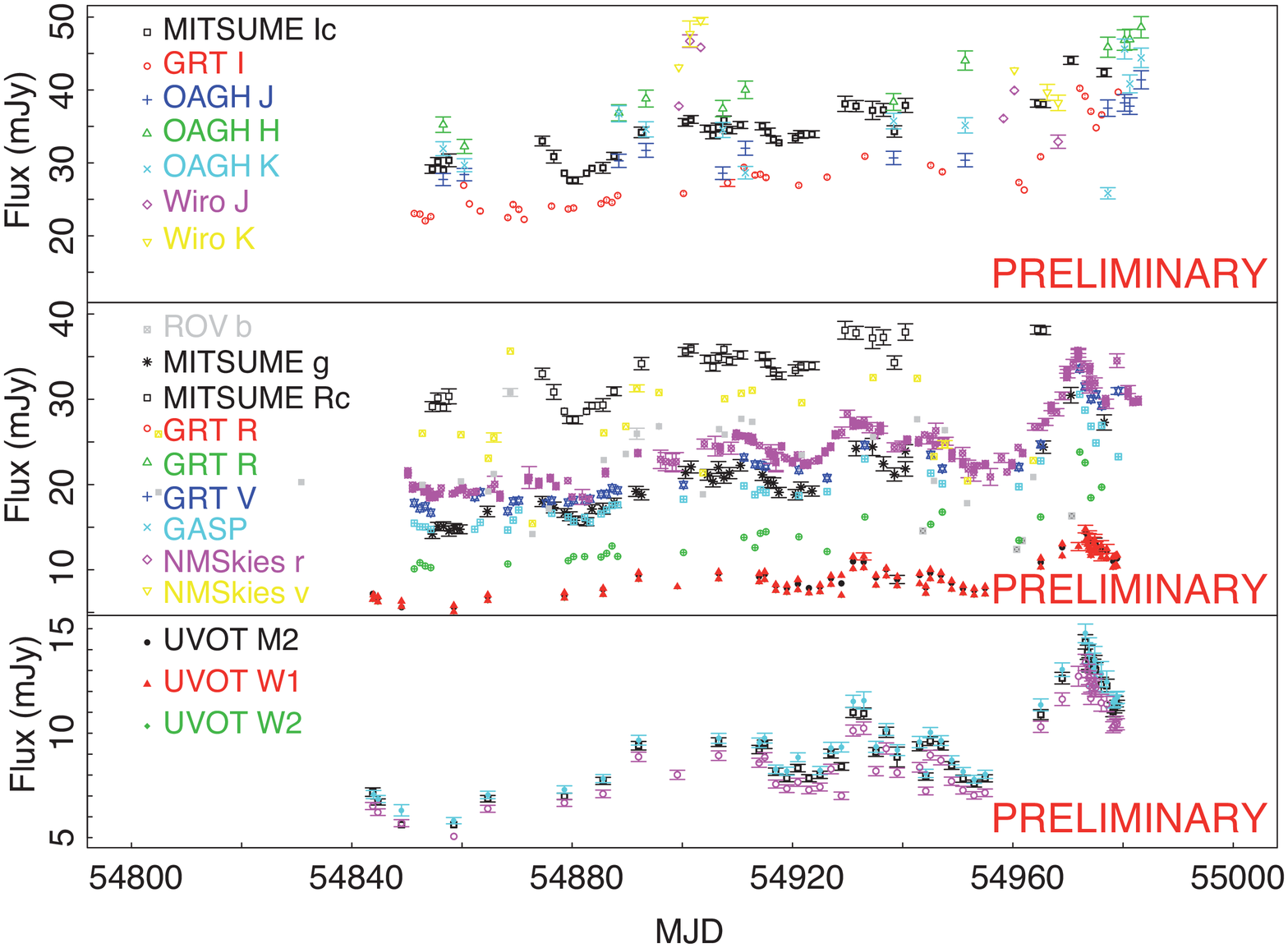}
\caption[]
        {{\it Top to bottom panel:} VHE (MAGIC, Whipple), Fermi-LAT, X-rays (RXTE and Swift) and 
          optical lightcurves of Mkn 421 between MJDs 54800 and 55000.
        }
\label{fig:Mkn421_lc}
\end{center}
\end{figure}

In the case of Mkn 421, no significant flaring activity was seen during the campaign, although some level of variability
was present throughout the monitoring period, with larger amplitudes at higher energies. As shown in Figure 
\ref{fig:Mkn421_lc}, the source was quiet for the entire observation period in all observational bands. 
The largest variability amplitude was registered in the X-rays where variations by a factor of two in flux were 
seen -- still much lower than the maximum values historically registered of 10-20 times variations. 



\subsection{Characterisation of the Variability}

\begin{figure}[!t]
  \vspace{5mm}
  \centering
  \includegraphics[width=3.in]{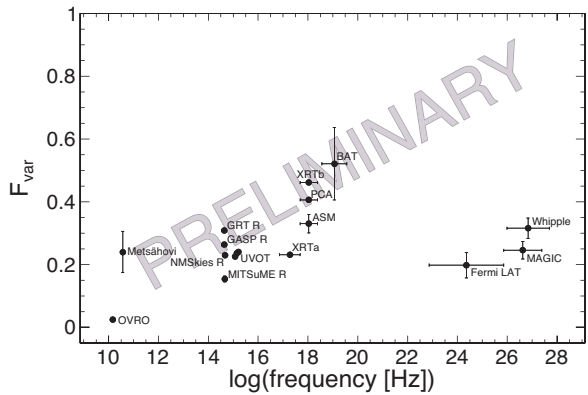}
  \caption{Fractional variability of Mrk 421 for all instruments taking part in the campaign at different wavebands.}
  \label{fig:Fvar421}
 \end{figure}

\begin{figure}[!t]
  \vspace{5mm}
  \centering
  \includegraphics[width=3.in]{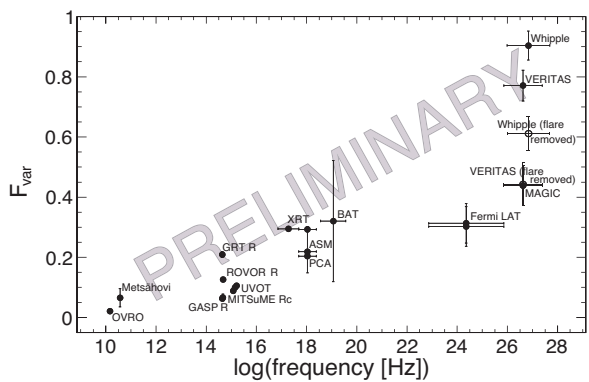}
  \caption{Fractional variability of Mrk 501 for all instruments taking part in the campaign at different wavebands.}
  \label{fig:Fvar501}
 \end{figure}

In order to quantify and characterise the variability at different energy bands, we have used the prescription for 
calculation of the fractional variability of the lightcurves, $F_{\rm{var}}$, given in \cite{Vaughan03}.
Figures \ref{fig:Fvar421} and \ref{fig:Fvar501} present the fractional variability plots of Mrk 421 and Mrk 501 for 
selected instruments at different bands across the SED. The plots show that even in the 
quiescent state, some level of variability is present for both sources, most significantly at the high energies 
(X-rays and VHE), which are usually associated to emission from the faster-cooling, most-energetic electrons, either at 
the synchrotron or the inverse-Compton channels.

For the case of Mkn 501 the variability of the VHE emission is dominant, even when the TeV flare observed by 
Whipple and VERITAS is exluded from the dataset ($F_{\rm{var}} \simeq 0.6$); in this case, 
MAGIC and VERITAS (when the flare is removed) measurements
show the same range of $F_{\rm{var}}$, even though observations are not strictly simultaneous. X-rays and Fermi-LAT GeV 
data show little variability
($F_{\rm{var}} \simeq 0.3$), and for the latter these are dominated by longer-term (30-day timescale) variations in flux.

It is interesting to notice the different behaviour shown in the case of Mkn 421, where the variability is dominated by a 
larger-amplitude $F_{\rm{var}}$ of the X-ray emission ($F_{\rm{var}} \simeq 0.5$). Despite the similarities shared by the 
electron populations of both sources (and the jet's physical parameters), as estimated from the SED modeling reported in 
\cite{Fermi11a, Fermi11b}, 
the fractional variability points to some differences between the source's behaviour when we look at their lightcurves
measured during the campaigns. This points to the necessity of time-dependent studies of these objects, which 
can reveal aspects of their physics not apparent from the time-averaged SEDs, hence suggesting that maybe 
other properties of the two systems differ, like for example the energy of the seed photon fields. 

The expected faster cooling times of the more energetic particles could naturally be understood 
as the responsible for the largest variability levels seen in X-rays and the VHE gamma-rays. It is also important to note 
that the sampling of the lightcurve by the different instruments was not homogeneous, a factor which can affect the 
variability estimations and will be closely investigated in a forthcoming publication.

\section{Conclusions and Outlook}

In this paper a brief summary of the results of two unprecedented MWL observational campaigns on the 
prototytpical TeV BL Lacs, Mkn 421 and Mkn 501, conducted in 2009, is presented. Lightcurves and variability studies are 
shown here for the 
first time. These \emph{preliminary} results show both sources to be in a relatively low state for most of the campaign 
(exception made to a flare of Mkn 501 observed in May 2009), thus allowing us to study with detail the quiescent state of 
both these objects. Despite their low state, both sources showed some level of variability, which is characteristically 
larger at higher energies, as expected from the faster cooling times of the most energetic electrons.  A deeper 
investigation of both sources will be a task for a future paper in preparation.

\section{Acknowledgments}
The MAGIC collaboration thanks the Instituto de Astrofisica de Canarias for the excellent working conditions at the Observatorio del Roque de los Muchachos in La Palma. The support of the German BMBF and MPG, the italian INFN and Spanish MICINN is gratefully acknoweledged. This work was also supported by the Marie Curie program, by the CPAN CSD2007-00042 and MultiDark CSD2009-00064 projects of the Spanish Consolider-Ingenio 2010 programme, by grant D002-353 of the Bulgarian NSF, by grant 127740 of the Academy of Finland, by the YIP of the Helmholtz Gemeinschaft, by the DFG Cluster of Excellence ``Origin and Structure of the Universe'', by the DFG Collaborative Research Centres SFB823/C4 and SFB876/C3, and by the Polish MNiSzW grant 745/N-HESS-MAGIC/2010/0. The VERITAS collaboration acknowledges support from the U.S. Department of Energy, the U.S. National Science Foundation and the Smithsonian Institution, by NSERC in Canada, by Science Foundation Ireland, and by STFC in the UK. We acknowledge the excellent work of the technical support at the FLWO and the collaboration institutions in the construction and operation of the instrument. The Fermi/LAT collaboration acknowledges support from a number of agencies and institutes for both development and the operation of the LAT as well as sicnetific data analysis. These include NASA and DOE in the United States, CEA Irfu and IN2P3 CNRS in France, ASI and INFN in Italy, MEXT, KEK, and JAXA in Japan, and the K. A. Wallenberg Foundation, the Swedish Research Council and the National Space Board in Sweden. Additional support from INAF in Italy and CNES in France for science analysis during the operations phase is acknowledged.


\clearpage


\begin{thebibliography}{}

\bibitem{Fermi11a} A.A. Abdo et al., 2011, ApJ, {\bf 727}(2): 129

\bibitem{Fermi11b} A.A. Abdo et al., 2011, ApJ, {\bf 736}(2): 131

\bibitem{MAGIC07} J. Albert et al.,2007, ApJ, {\bf 669}(2): 862

\bibitem{MAGIC10a} J. Aleksi\'{c} et al., 2010, A\&A, {\bf 519}: 32

\bibitem{MAGIC09a} H. Anderhub et al., 2009, ApJ, {\bf 705}(2): 1624.

\bibitem{Fossati08} G. Fossati, J.H. Buckley et al., 2008, ApJ, {\bf 677}: 906-25 

\bibitem{Katarzynski10} H. Katarzy\'{n}ski, 2010, A\&A, {\bf 510}: 63

\bibitem{Mastichiadis08} A. Mastichiadis \& K. Moraitis, 2008, A\&A, {\bf 491}: L37 


\bibitem{Pichet} M. Pichel and D. Paneque, 2011, these proceedings

\bibitem{Punch92} M. Punch et al., 1992, Nature, {\bf 358}(6386): 477.

\bibitem{Quinn96} J. Quinn et al., 1996, ApJL, {\bf 456}: L83

\bibitem{Tluczykont10} M. Tluczykont et al., 2010, A\&A, {\bf 524}: 48

\bibitem{Vaughan03} S. Vaughan et al. 2003, MNRAS, {\bf 345}: 1271

\bibitem{Zweerink97} J.A. Zweerink et al., 1997, ApJL, {\bf 490}: L141

\end{thebibliography}
\end{document}